\documentclass[letterpaper,twocolumn,10pt]{article}
\usepackage{usenix2019_v3,epsfig}

\usepackage{tikz}
\usepackage{amsmath}

\usepackage{filecontents}

\usepackage{comment}
\usepackage{algorithm,algpseudocode}
\usepackage{url}            

\usepackage{booktabs}   
\usepackage{subcaption} 

\usepackage{color}
\definecolor{dkgreen}{rgb}{0,0.6,0}
\definecolor{gray}{rgb}{0.5,0.5,0.5}
\definecolor{mauve}{rgb}{0.58,0,0.82}

\usepackage{pdfpages}
\usepackage{balance}

\usepackage{xcolor}

\usepackage{multirow}
\usepackage{amssymb}
\usepackage{pifont}
\usepackage{soul}
\newcommand{\cmark}{\ding{51}}%
\newcommand{\xmark}{\ding{55}}%
\newcommand\blfootnote[1]{%
  \begingroup
  \renewcommand\thefootnote{}\footnote{#1}%
  \addtocounter{footnote}{-1}%
  \endgroup
}

\begin{document}
\linespread{0.95}

\date{}

\title{\Large \bf Optimizing CNN Model Inference on CPUs}
\author{Yizhi Liu\textsuperscript{*}, Yao Wang\textsuperscript{*}, Ruofei Yu, Mu Li, Vin Sharma, Yida Wang \\ Amazon Web Services \\ \{yizhiliu, wayao, yuruofei, mli, vinarm, wangyida\}@amazon.com}
\maketitle

\pagenumbering{gobble}

\subsection*{Abstract}
The popularity of Convolutional Neural Network (CNN) models and the ubiquity of CPUs imply that better performance of CNN model inference on CPUs can deliver significant gain to a large number of users. To improve the performance of CNN inference on CPUs, current approaches like MXNet and Intel OpenVINO usually treat the model as a graph and use the high-performance libraries such as Intel MKL-DNN to implement the operations of the graph. While achieving reasonable performance on individual operations from the off-the-shelf libraries, this solution makes it inflexible to conduct optimizations at the graph level, as the local operation-level optimizations are predefined. Therefore, it is restrictive and misses the opportunity to optimize the end-to-end inference pipeline as a whole. This paper presents \emph{NeoCPU}, a comprehensive approach of CNN model inference on CPUs that employs a full-stack and systematic scheme of optimizations. \emph{NeoCPU} optimizes the operations as templates without relying on third-parties libraries, which enables further improvement of the performance via operation- and graph-level joint optimization. Experiments show that \emph{NeoCPU} achieves up to 3.45$\times$ lower latency for CNN model inference than the current state-of-the-art implementations on various kinds of popular CPUs. \blfootnote{\textsuperscript{*}Equal contribution}

\section{Introduction}
\label{sec:intro}
The growing use of Convolutional Neural Network (CNN) models in computer vision applications makes this model architecture a natural focus for performance optimization efforts. Similarly, the widespread deployment of CPUs in servers, clients, and edge devices makes this hardware platform an attractive target. Therefore, performing CNN model inference efficiently on CPUs is of critical interest to many users.

The performance of CNN model inference on CPUs leaves significant room for improvement. Performing a CNN model inference is essentially executing a computation graph consisting of operations. In practice, people normally use high-performance kernel libraries (e.g. Intel MKL-DNN~\cite{mkldnn} and OpenBlas~\cite{xianyi2014openblas}) to obtain high performance for CNN operations. While these libraries tune very carefully for common operations with normal input data shapes (e.g. 2D convolutions), they only focus on the (mostly, convolution) operations but miss the opportunities to further optimize the end-to-end model inference at the graph level. The graph-level optimization is often handled by the deep learning frameworks, e.g. TensorFlow~\cite{abadi2016tensorflow} and MXNet~\cite{chen2015mxnet}.

However, the graph-level optimization such as operation fusion and data layout planing that a framework can do is limited because the operation implementation is already predefined in the third-party libraries. Therefore, the optimizations in the frameworks do not work in concert with the optimizations in the kernel library, which leaves significant performance gains unrealized in practice. Furthermore, different CPU architectures rely on different high-performance libraries and integrating a library into a deep learning framework requires error-prone and time-consuming engineering effort. Lastly, although those libraries are highly optimized, they present as third-party plug-ins, which may introduce contention issues with other libraries in the framework. As an example, TensorFlow originally used the Eigen library~\cite{eigen} to handle computation on CPUs. Later on, MKL-DNN was also introduced. As a consequence, at runtime MKL-DNN threads coexist with Eigen threads, resulting in resource contention. In summary, this kind of \emph{framework-specific} approach for CNN model inference on CPUs is inflexible, cumbersome, and sub-optimal.

Because of the constraint imposed by the framework, optimizing the performance of CNN model inference end-to-end without involving a framework (i.e. a \emph{framework-agnostic} method) is of obvious interest to many deep learning practitioners. Recently, Intel launched a universal CNN model inference engine called OpenVINO Toolkit~\cite{openvino}. This toolkit optimizes CNN models in the computer vision domain on Intel processors (mostly x86 CPUs) and claims to achieve better performance than the deep learning frameworks alone. Yet, OpenVINO could only provide limited graph-level optimization (e.g. operation fusion as implemented in ngraph~\cite{cyphers2018intel}) as it still relies upon MKL-DNN to deliver performance gains for the carefully-tuned operations. Therefore, the optimization done by OpenVINO is still not sufficient for most of the CNN models.

Based on the previous observation, we argue that in order to further improve the CNN model inference performance on CPUs, being able to do the \emph{flexible end-to-end optimization} is the key. In this paper, we propose \emph{NeoCPU}, a comprehensive approach to optimize CNN models for efficient inference on CPUs. \emph{NeoCPU} is full-stack and systematic, which includes operation- and graph-level joint optimizations and does not rely on any third-party high-performance libraries. At the operation level, we follow the well-studied techniques to optimize the most computationally-intensive operations like convolution (\emph{CONV}) in a \emph{template}, which is applicable to different workloads on multiple CPU architectures and enables us for flexible graph-level optimization. At the graph level, in addition to the common techniques such as operation fusion and inference simplification, we coordinate the individual operation optimizations by manipulating the data layout flowing through the entire model for the best end-to-end performance. In summary, \emph{NeoCPU} does the end-to-end optimization in a flexible and automatic fashion, while the existing works rely on third-party libraries and lack comprehensive performance tuning.

\emph{NeoCPU} is built upon a deep learning compiler stack named TVM~\cite{chen2018tvm} with a number of enhancements. TVM enables the possibility of using own operation-level optimizations instead of third-party high-performance libraries, which make it flexible to apply our operation- and graph-level joint optimization. However, there was only one customized operation-level optimization on ARM CPUs for convolutions with specific data shapes and no operation- and graph-level joint optimization in the original TVM stack before our work. In addition, there exist other deep learning compilers such as Tensor Comprehensions~\cite{vasilache2018tensor} and Glow~\cite{rotem2018glow}. Unfortunately, they either do not target on CPUs or not optimize the CPU performance well, e.g. based on the paper description and our own experiments, Glow only optimizes the single-core performance for CPUs. Therefore we do not incorporate those works as the baseline. Table~\ref{tbl:overview} summarizes the features of \emph{NeoCPU} compared to others. To the best of our knowledge, \emph{NeoCPU} achieves competitive performance for CNN model inference on various kinds of popular CPUs.

\begin{table}[tbp]
\centering
	\resizebox{0.48\textwidth}{!}{%
    \begin{tabular}{ccccc}
    \midrule
                     & \textbf{Op-level opt} & \textbf{Graph-level opt} & \textbf{Joint opt} & \textbf{Open-source} \\
                     \hline\\[-1.75ex]
    NeoCPU     & \cmark & \cmark & \cmark & \cmark \\
    MXNet\cite{chen2015mxnet}/TensorFlow\cite{abadi2016tensorflow} & 3rd party & limited & \xmark & \cmark \\
    OpenVINO\cite{openvino}         & 3rd party & limited & \textbf{?} & \xmark \\
    Original TVM\cite{chen2018tvm}  & incomplete & \cmark & \xmark & \cmark \\
    Glow\cite{rotem2018glow}             & single core & \cmark & \xmark & \cmark \\
    \midrule
    \end{tabular}
    }
    \caption{Side-by-side comparison between \emph{NeoCPU} and existing works on CNN model inference}
    \vspace{-1em}
\label{tbl:overview}
\end{table}

Specifically, this paper makes the following contributions:
\begin{itemize}
\item Provides an operation- and graph-level joint optimization scheme to obtain high CNN model inference performance on different popular CPUs including Intel, AMD and ARM, which outperforms the current state-of-the-art implementations;
\item Constructs a template to achieve good performance of convolutions, which is flexible to apply to various convolution workloads on multiple CPU architectures (x86 and ARM) without relying on high-performance kernel libraries;
\item Designs a global scheme to look for the best layout combination in different operations of a CNN model, which minimizes the data layout transformation overhead between operations while maintaining the high performance of individual operations.
\end{itemize}

It is worth noting that, this paper primarily deals with direct convolution computation, while \emph{NeoCPU} is compatible to other optimziation works on the computationally-intensive kernels, e.g. \emph{CONV}s via Winograd~\cite{budden2016deep,jia2018winograd} or FFT~\cite{zlateski2018fft}.

We evaluated \emph{NeoCPU} on CPUs with both x86 and ARM architectures. In general, \emph{NeoCPU} delivers the best performance for 13 out of 15 popular networks on Intel Skylake CPUs, 14 out of 15 on AMD EYPC CPUs, and all 15 models on ARM Cortex A72 CPUs. It is worthwhile noting that the baselines on x86 CPUs were more carefully tuned by the chip vendor (Intel MKL-DNN) but the ARM CPUs were less optimized. While the selected framework-specific (MXNet and TensorFlow) and framework-agnostic (OpenVINO) solutions may perform well on one case and less favorably on the other case, \emph{NeoCPU} runs efficiently across models on different architectures.

In addition, \emph{NeoCPU} produces a standalone module with minimal size that does not depend on either the frameworks or the high-performance kernel libraries, which enables easy deployment to multiple platforms. \emph{NeoCPU} is used in Amazon SageMaker Neo Service~\footnote{\url{https://aws.amazon.com/sagemaker/neo/}}, enabling model developers to optimize for inference on CPU-based servers in the cloud and devices at the edge. Using this service, a number of application developers have deployed CNN models optimized for inference in production on several types of platforms. All source code has been released to the open source TVM project\footnote{\url{https://github.com/dmlc/tvm}}.

The rest of this paper is organized as follows: Section~\ref{sec:back} reviews the background of modern CPUs as well as the typical CNN models; Section~\ref{sec:opt} elaborates the optimization ideas that we propose and how we implement them, followed by evaluations in Section~\ref{sec:eval}. We list the related works in Section~\ref{sec:related} and summarize the paper in Section~\ref{sec:concl}.

\section{Background}
\label{sec:back}
\subsection{Modern CPUs}
\label{sec:back:cpu}
Although accelerators like GPUs and TPUs demonstrate their outstanding performance on the deep learning workloads, in practice, there is still a significant number of deep learning computation, especially model inference, taking place on the general-purpose CPUs due to the high availability. Currently, most of the CPUs equipped on PCs and servers are manufactured by Intel or AMD with x86 architecture~\cite{cpumarketshare}, while ARM CPUs with ARM architecture occupy the majority of embedded/mobile device market~\cite{armmarketshare}.

Modern CPUs use thread-level parallelism via multi-core~\cite{hammond2000stanford} to improve the overall processor performance given the diminishing increasing of transistor budgets to build larger and more complex uniprocessor. It is critical to avoid the interference among threads running on the same processor and minimize their synchronization cost in order to have good scalability on multi-core processors. Within the processor, a single physical core achieves the peak performance via the SIMD (single-instruction-multiple-data) technique. SIMD loads multiple values into wide vector registers to process together. For example, Intel introduced the 512-bit Advanced Vector Extension instruction set (AVX-512), which handles up to 16 32-bit single precision floating point numbers (totally 512 bits) per CPU cycle. And the less advanced AVX2 processes data in 256-bit registers. In addition, these instruction sets utilize the Fused-Multiply-Add (FMA) technique which executes one vectorized multiplication and then accumulates the results to another vector register in the same CPU cycle. The similar SIMD technique is embodied in ARM CPUs as NEON~\cite{armneon}. As shown in the experiments, our proposed solution works on both x86 and ARM architectures.

In addition, it is worth noting that modern server-side CPUs normally supports hyper-threading~\cite{marr2002hyper} via the simultaneous multithreading (SMT) technique, in which the system could assign two virtual cores (i.e. two threads) to one physical core, aiming at improving the system throughput. However, the performance improvement of hyper-threading is application-dependent~\cite{leng2002empirical}. In our case, we do not use hyper-threading since one thread has fully utilized its physical core resource and adding one more thread to the same physical core will normally decrease the performance due to the additional context switch. We also restrict our optimization within processors using the shared-memory programming model as this is the typical system setting for CNN model inference. The
Non-Uniformed Memory Access (NUMA) pattern occurred
in the context of multiple processors on the same motherboard
is beyond the scope of this paper.

\subsection{Convolutional neural networks}
\label{sec:back:cnn}
Convolutional neural networks (CNNs) are commonly used in computer vision workloads~\cite{krizhevsky2012imagenet,simonyan2014very,szegedy2015going,szegedy2016rethinking,he2016deep,huang2017densely,liu2016ssd}. A CNN model is normally abstracted as a computation graph, essentially, Directed Acyclic Graph (DAG), in which a node represents an operation and a directed edge pointing from node X to Y represents that the output of operation X serves as (a part of) the inputs of operation Y (i.e. Y cannot be executed before X). Executing a model inference is actually to flow the input data through the graph to get the output. Doing the optimization on the graph (e.g. prune unnecessary nodes and edges, pre-compute values independent to input data) could potentially boost the model inference performance.

Most of the computation in the CNN model inference attributes to \emph{convolutions (CONVs)}. These operations are essentially a series of multiplication and accumulation, which by design can fully utilize the parallelization, vectorization and FMA features of modern CPUs. Existing works~\cite{mkldnn,heinecke2016libxsmm,georganas2018anatomy} have demonstrated that it is possible to achieve high performance of convolution operations on CPUs by arranging the data layout and consequently, the computation, in an architecture-friendly way. The remaining challenge is how to manage the data layout flowing through these operations efficiently to get the high performance out of the end-to-end CNN model inference.

The rest of the CNN workloads are mostly memory-bound operations associated to \emph{CONV}s (e.g. batch normalization, pooling, activation, element-wise addition, etc.). The common practice~\cite{chen2018tvm} is fusing them to \emph{CONV}s so as to increase the overall arithmetic intensity of the workload and consequently boost the performance.

The computation graph of CNN model training has no essential difference with inference, just being larger (adding in backwards operations) and with some more computationally-trivial operations (e.g. loss function). Therefore, the optimization work done for CNN model inference is applicable to training as well.
\section{Optimizations}
\label{sec:opt}
This section describes our optimization ideas and implementations in detail. The solution presented in this paper is end-to-end for doing the CNN model inference. Our proposed solution is generic enough to work for a wide range of common CNN models as we will show in the evaluation. The basic idea of our approach is to view the optimization as an end-to-end problem and search for a globally best optimization. That is, we are not biased towards a local performance optimal of a single operation as many previous works. In order to achieve this, we first present how we optimized the computationally intensive convolution operations at low-level using a configurable template (Section~\ref{sec:opt:op}). This makes it flexible to search for the best implementation of a specific convolution workload on a particular CPU architecture, and to optimize the entire computation graph by choosing proper data layouts between operations to eliminate unnecessary data layout transformation overhead (presented in Section~\ref{sec:opt:graph} and~\ref{sec:opt:search}).

We implemented the optimization based on the TVM stack~\cite{chen2018tvm} by adding a number of new features to the compiling pass, operation scheduling and runtime components. The original TVM stack has done a couple of generic graph-level optimizations including operation fusion, pre-computing, simplifying inference for batch-norm and dropout~\cite{chen2018tvm}, which are also inherited to this work but will not be covered in this paper.

\subsection{Operation optimization}
\label{sec:opt:op}
Optimizing convolution operations is critical to the overall performance of a CNN workload as it takes the majority of computation. This is a well-studied problem but the previous works normally go deep to the assembly code level for high performance~\cite{mkldnn,heinecke2016libxsmm}. In this subsection, we show how to take advantage of the latest CPU features (SIMD, FMA, parallelization, etc.) to optimize a single \emph{CONV} without going into the tedious assembly code or C++ intrinsics. By managing the implementation in high-level instead, it is then easy to extend our optimization from a single operation to the entire computation graph.

\subsubsection{Single thread optimization}
\label{sec:opt:op:memory}
\begin{figure}[htbp]
\centering
\includegraphics[width=0.5\textwidth]{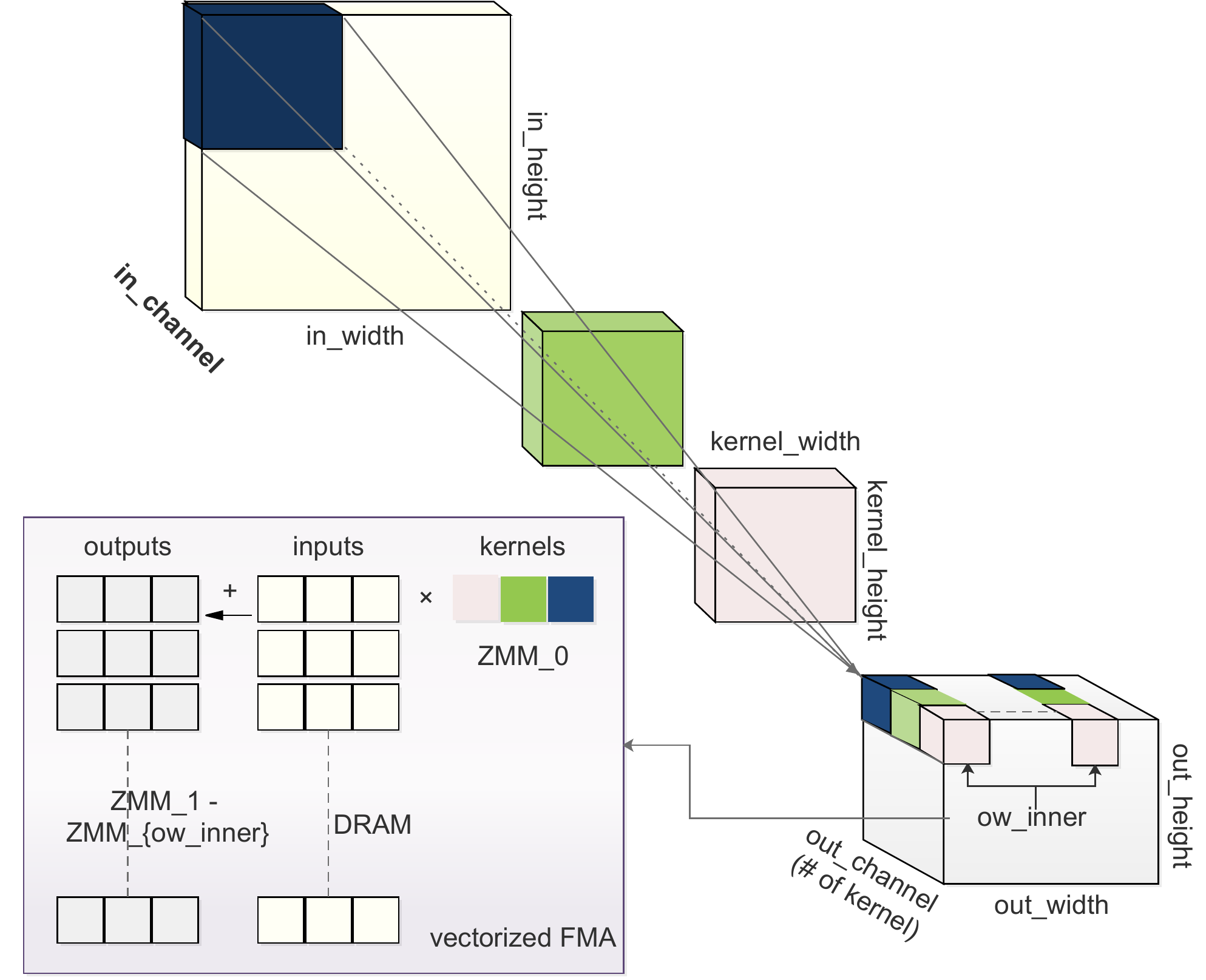}
\caption{The illustration of \emph{CONV} and the efficient implementation in AVX-512 instructions as an example. There are three kernels depicted in dark blue, green and light pink. To do efficient FMA, multiple kernel values are packed into one $ZMM$ register and reused to multiply with different input values and accumulate to output values in different $ZMM$ registers.}
\label{fig:conv}
\end{figure}

We started from optimizing \emph{CONV} within one thread. \emph{CONV} is computationally-intensive which traverses its operands multiple times for computation. Therefore, it is critical to manage the layout of the data fed to the \emph{CONV} to reduce the memory access overhead. We first revisit the computation of \emph{CONV} to illustrate our memory management scheme. A 2D \emph{CONV} in CNN takes a 3D feature map (height $\times$ width $\times$ channels) and a number of 3D convolution kernels (normally smaller height and width but the same number of channels) to convolve to output another 3D tensor. The calculation is illustrated in Figure~\ref{fig:conv}, which implies loops of 6 dimensions: \emph{in\_channel}, \emph{kernel\_height}, \emph{kernel\_width}, \emph{out\_channel}, \emph{out\_height} and \emph{out\_width}. Each kernel slides over the input feature map along the height and width dimensions, does element-wise product and accumulates the values to produce the corresponding element in the output feature map, which can naturally leverage FMA. The number of kernels forms \emph{out\_channel}. Note that three of the dimensions (\emph{in\_channel},  \emph{kernel\_height} and \emph{kernel\_width}) are reduction axes that cannot be embarrassingly parallelized.

We use the conventional notation \emph{NCHW} to describe the default data layout, which means the input and output are 4-D tensors with \emph{batch size} N, \emph{number of channels} C, \emph{feature map height} H, \emph{feature map width} W, where \emph{N} is the outermost and \emph{W} is the innermost dimension of the data. The related layout of kernel is \emph{KCRS}, in which \emph{K}, \emph{C}, \emph{R}, \emph{S} stand for the output channel, input channel, kernel height and kernel width.

Following the common practice~\cite{tf-NHWC, mkldnn}, we organized the feature map layout as \emph{NCHW[x]c} for better memory access patterns i.e. better cache locality, in which \emph{c} is a split sub-dimension of channel \emph{C} in super-dimension, and the number \emph{x} indicates the split size of the sub-dimension (i.e. $\#channels = sizeof(C) \times sizeof(c)$, where $sizeof(c)=x$). The output has the same layout \emph{NCHW[y]c} as the input, while the split factor can be different. Correspondingly, the convolution kernel is organized in \emph{KCRS[x]c[y]k}, in which \emph{c} with split size $x$ and \emph{k} with split size $y$ are the sub-dimensions of input channel \emph{C} and output channel \emph{K}, respectively. It is worth noting that a significant amount of data transformation overhead needs to be paid to get the desired layout.

In addition to the dimension reordering, for better utilizing the latest vectorization instructions (e.g. AVX-512, AVX2, NEON, etc.), we split \emph{out\_width} to \emph{ow\_outer} and \emph{ow\_inner} using a factor \emph{reg\_n} and move the loop of \emph{ow\_inner} inside for register blocking. For example, on a CPU featured AVX-512, we can utilize its 32 512-bit width registers $ZMM_0 - ZMM_{31}$~\cite{avx512ins} as follows. We maintain the loop hierarchy to use one ZMM register to store the kernel data while others storing the feature map. The kernel values stored in one $ZMM$ register (up to 512 bits, a.k.a, 16 output channels in float32) are used to multiply with a number of input feature map values continuously stored in the DRAM via AVX-512F instructions~\cite{avx512ins}, whose results are then accumulated to other $ZMM$ registers storing the output values. Figure~\ref{fig:conv} illustrates this idea. For other vectorized instructions, the same idea applies but the split factor of \emph{out\_width} (i.e. \emph{reg\_n}) may change.

Algorithm~\ref{alg:conv} summarizes our optimization of \emph{CONV} in single thread, which essentially is about 1) dimension ordering for friendly memory locality and 2) register blocking for good vectorization instruction utilization, as in previous works. However, unlike others, we made it a \emph{template} in high-level language
, in which the block size ($x$, $y$), the number of utilized registers (\emph{reg\_n}), and the loop-unroll strategy ($unroll\_ker$) are easily configurable. Consequently, the computing logic can be adjusted according to different CPU architectures (cache size, registered vector width, etc.) as well as different workloads (feature map size, convolution kernel size, etc.). This is flexible and enables graph-level optimization we will discuss later.

\begin{algorithm}[htbp]
    \caption{CONV operation algorithm via FMA}
    \label{alg:conv}
    \begin{algorithmic}[1]
        \State PARAM: $x > 0$ s.t. $in\_channel \bmod{x} = 0$
        \State PARAM: $y > 0$ s.t. $out\_channel \bmod{y} = 0$
        \State PARAM: $reg\_n > 0$ s.t. $out\_width \bmod{reg\_n} = 0$
        \State PARAM: $unroll\_ker \in \{True, False\}$
        \State INPUT: $IFMAP$ in NCHW[x]c
        \State INPUT: $KERNEL$ in KCRS[x]c[y]k
        \State OUTPUT: $OFMAP$ in NCHW[y]c
        \For {each disjoint chunk of $OFMAP$} \Comment{parallel}
        \For {ow.outer:= $0 \to out\_width / reg\_n$}
        \State Initialize $V\_REG_1$ to $V\_REG_{reg\_n}$ by $\vec{0}$
        \For {ic.outer:= $0 \to in\_channel / x$}
        \For {each entry of $KERNEL$} \Comment{(opt) unroll}
        \For {ic.inner:= $0 \to x$}
        \State $vload(KERNEL, V\_REG_0)$ \Comment{y floats}
        \For {i:= $1 \to reg\_n+1$} \Comment{unroll}
        \State $vfmadd(IFMAP, V\_REG_0, V\_REG_i)$
        \EndFor
        \EndFor
        \EndFor
        \EndFor
        \For {i:= $1 \to reg\_n+1$}
        \State $vstore(V\_REG_{i}, OFMAP)$
        \EndFor
        \EndFor
        \EndFor
    \end{algorithmic}
\end{algorithm}

\subsubsection{Thread-level parallelization}
\label{sec:opt:op:parallel}
It is a common practice to partition \emph{CONV} into disjoint pieces to parallelize among multiple cores of a modern CPU. Kernel libraries like Intel MKL-DNN usually uses off-the-shelf multi-threading solution such as OpenMP. However, we observe that the resulting scalability of the off-the-shelf parallelization solution is not desirable (Section~\ref{sec:eval:scale}).

Therefore, we implemented a customized thread pool to efficiently process this kind of embarrassing parallelization. Basically, in a system of $N$ physical cores, we evenly divided the outermost loop of the operation into $N$ pieces to assign to $N$ threads. Then we used C++11 atomics to coordinate threads during fork-join and an single-producer-single-consumer lock-free queue between the scheduler and every working thread to assign tasks. Active threads are guaranteed to run on disjoint physical cores via thread binding to minimize the hardware contention, and no hyper-threading is used as discussed in Section~\ref{sec:back:cpu}. For the global data structure accessed by multiple threads such as the lock-free queues, we inserted cache line padding as needed to avoid false sharing between threads. In summary, this customized thread pool employs deliberate mechanism to prevent resource contention and reduce the thread launching overhead, which makes it outperform OpenMP according to our evaluation.

\subsection{Layout transformation elimination}
\label{sec:opt:graph}
In this subsection, we extend the optimization scope from a single operation to the entire computation graph of the CNN model. The main idea here is to come up with a generic solution at the graph level to minimize the data layout transformation introduced by the optimization in Section~\ref{sec:opt:op}. Previous works~\cite{mkldnn,heinecke2016libxsmm,georganas2018anatomy} which focus on individual operation optimization normally do not consider about the data layout transformation overhead between highly optimized operations.

Since \emph{NCHW[x]c} is efficient for \emph{CONV}s which takes the majority of the CNN model computation, we should make sure that every \emph{CONV} is executed in this layout. However, other operations between \emph{CONV}s may only be compatible with the default layout, which makes each \emph{CONV} transform the input data layout from default (\emph{NCHW} or \emph{NHWC}) to \emph{NCHW[x]c} before the computation and transform it back at the end. This transformation introduces significant overhead.

Fortunately, from the perspective of the graph level, we can take the layout transformation out of \emph{CONV} to be an independent node, and insert it only when necessary. That is, we eliminate the transformation taking place in the \emph{CONV} operation and maintain the transformed layout flow through the graph as far as possible.

\begin{figure*}[!htbp]
\centering
\includegraphics[width=0.7\textwidth]{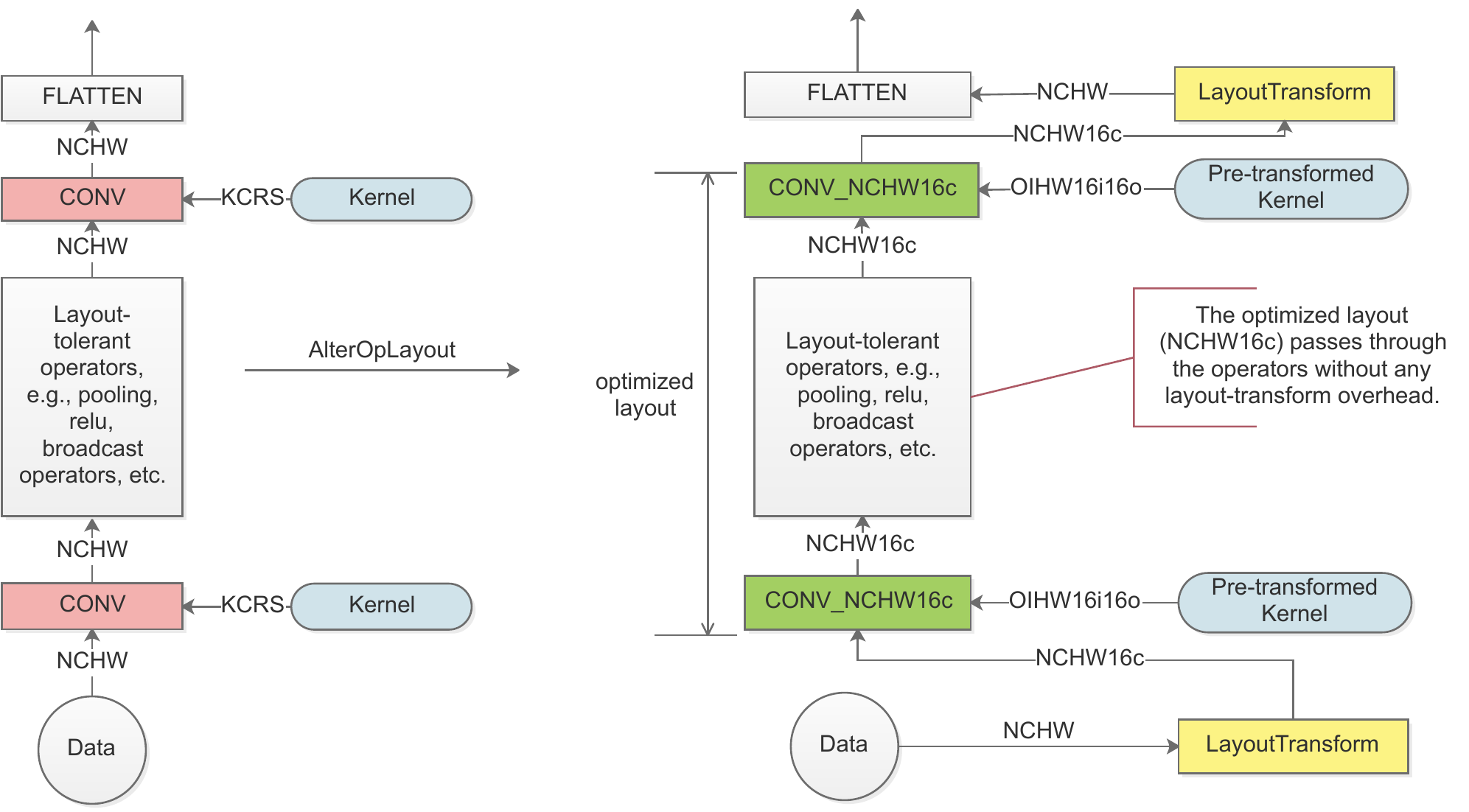}
\caption{Layout optimization of a simple CNN model. The notation on an edge represents the layout of the data passing through this edge. The left side depicts the network with default data layout. Each \emph{CONV} node in pink needs to pay additional overhead to transform the data into a favorable layout to achieve good performance and then transform back to default. The network in the right side is optimized at the graph level to minimize the data layout transformation during the runtime. The \emph{CONV} nodes in green do not need to transform any data before and after computation.}
\label{fig:layout}
\end{figure*}

In order to determine if a data transformation is necessary, we first classify operations into three categories according to how they interact with the data layout as follows:

\begin{enumerate}
\itemsep0em 
\item \emph{Layout-oblivious} operations. These operations process the data without the knowledge of its layout, i.e. it can handle data in any layout. Unary operations like \emph{ReLU}, \emph{Softmax}, etc., fall in this category.
\item \emph{Layout-tolerant} operations. These operations need to know the data layout for processing, but can handle a number of layout options. For example, \emph{CONV}, in our case, can deal with \emph{NCHW}, \emph{NHWC} and \emph{NCHW[x]c} layouts. Other operations like \emph{Batch\_Norm}, \emph{Pooling}, etc., fall in this category as well.
\item \emph{Layout-dependent} operations. These operations process the data only in one specific layout, that is, they do not tolerate any data transformation. Therefore, the layout has to be transformed to a certain format before passing to a layout-dependent operation. Transformation operations like \emph{Flatten}, \emph{Reshape}, etc, fall in this category.
\end{enumerate}

Operations between \emph{CONV}s in typical CNN models are either layout-oblivious (e.g. \emph{ReLU}, \emph{SoftMax}, \emph{Concat}, and \emph{ElemwiseAdd}) or layout-tolerant (e.g. \emph{Batch\_Norm}, \emph{Pooling}), making it possible to keep the data layout being \emph{NCHW[x]c} across convolution layers. Layout transformation from \emph{NCHW} to \emph{NCHW[x]c} happens before the first \emph{CONV}. Data layout between \emph{CONV}s can be maintained the same (i.e. \emph{NCHW[x]c} sharing the same \emph{x} value) without transformation. Only if getting to a layout-dependent operation, e.g. \emph{Flatten}, the data layout is transformed back from \emph{NCHW[x]c} to \emph{NCHW}.

In practice, we first traverse the computation graph to infer the data layout of each node as illustrated in the left side of Figure~\ref{fig:layout}, then we alter the layout of \emph{CONV}s from default to \emph{NCHW[x]c} for better performance. Note that in order to prevent further transformation, we make $x$ a constant number (e.g. 16) across all \emph{CONV}s. However, this value may vary across different \emph{CONV}s in order to get the optimal performance, which requires layout transformation. We will explain more about this in Section~\ref{sec:opt:search}. Finally, the \emph{LayoutTransform} nodes are inserted to the graph accordingly. Thus, we still have \emph{NCHW} input and output for the network, but the internal layouts between \emph{CONV} layers are in optimized \emph{NCHW[x]c}, as shown in the right part of Figure~\ref{fig:layout}. It is worth noting that, the layout of the model parameters such as convolution kernel weights and the mean and variance of \emph{Batch\_Norm} are invariant so can be pre-transformed during the compilation. We also illustrate this in the right part of Figure~\ref{fig:layout}.

We implemented the ideas by introducing multiple graph-level optimization \emph{passes} to the TVM stack. By keeping transformed data layout invariant between \emph{CONV} layers as much as possible and pre-transforming the layout of convolution kernel weights at compilation time, we further improve the end-to-end performance of CNN model inference.

\subsection{Optimization scheme search}
\label{sec:opt:search}
We came up with the aforementioned optimization schemes, especially, how to layout the data, based on our understanding of the hardware, e.g. cache size, vectorization unit width, memory access pattern, etc. However, it is tedious and impractical to exhaust all possible optimal cases by hand. As a trade-off, Section~\ref{sec:opt:graph} assumes that the split factor of the channel, i.e. $x$ in \emph{NCHW[x]c}, stays the same during the entire network, while having various $x$ values in different \emph{CONV}s may lead to a better performance. In addition, the split factor of the output width, i.e. $reg\_n$, also needs to adjust for different vectorization instruction sets.

Therefore, an automatic search for the best scheme is in demand to further improve the performance. Basically, we should build a system to allow the domain experts to construct the search space for the machine to explore for the best scheme resulting in the shortest execution time. The search is two-stage, first local to find optimization scheme candidates for the individual computationally-intensive operations, then global to select and combine the individual schemes for the optimal end-to-end results. It is feasible to conduct this kind of search given the optimization template described in Section~\ref{sec:opt:op}. 

\begin{figure*}[!htbp]
\centering
\includegraphics[width=0.92\textwidth]{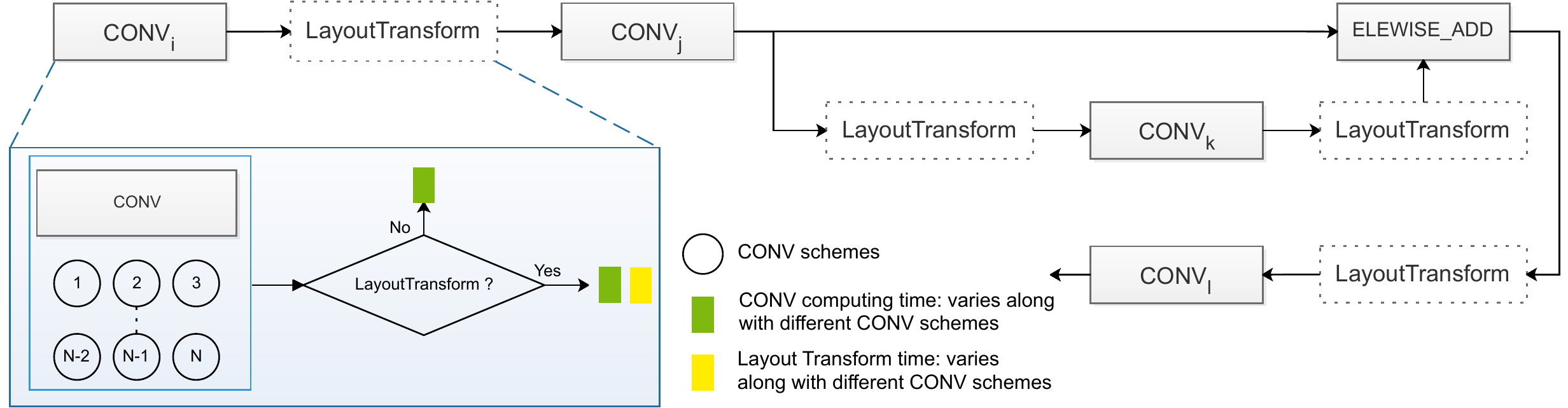}
\caption{Global search for CNN model inference. \emph{LayoutTransform} may or may not be in invoked according to the global decision. If invoked, an additional overhead of data transformation denoted in yellow needs to be paid.}
\label{fig:search}
\end{figure*}

\subsubsection{Local search}
\label{sec:opt:search:local}
The first step is to find the optimal schedules for each computationally-intensive operations, i.e. \emph{CONV}s in a CNN model. We used a tuple \emph{(ic\_bn, oc\_bn, reg\_n, unroll\_ker)} to represent a convolution schedule, whose items are chosen to cover different CPU architectures and generations for different convolution workloads. The first two terms \emph{ic\_bn} and \emph{oc\_bn} stand for the split factors of input and output channels (i.e. $x$ in the \emph{NCHW[x]c} notation), which are relevant to the cache sizes of a specific CPU. The third term \emph{reg\_n} is the number of SIMD registers to be used at the inner loop, which varies among different CPU architectures and generations. Also, we observed that utilizing all SIMD registers in a single thread does not always return the best performance. The last term \emph{unroll\_ker} is a boolean deciding whether to unroll the \emph{for} loop involving convolution kernel computation (line 12 of Algorithm~\ref{alg:conv}), as in some scenarios unrolling this loop may increase the performance by reducing branch penalties and such. The local search uses the template discussed in~\ref{sec:opt:op:memory} to find the best combination of these values to minimize the \emph{CONV} execution time, similar to the kernel optimization step in~\cite{jiang2018efficient}.

Specifically, the local search works as follows:

\begin{enumerate}
\itemsep0em 
\item Define the candidate lists of \emph{ic\_bn} and \emph{oc\_bn}. To exhaust the possible cases, we include all factors of the number of channels. For example, if the number of channels is 64, [32, 16, 8, 4, 2, 1] are listed as the candidates.
\item Define the candidate list of \emph{reg\_n}. In practice, we choose the \emph{reg\_n} value from [32, 16, 8, 4, 2].
\item Define the candidate list of \emph{unroll\_ker} to be [\emph{True}, \emph{False}].
\item Walk through the defined space to measure the execution time of all combinations, each of which will be run multiple times for averaging to cancel out the possible variance rooted from the unexpected interference from the operating system and/or other processes. This eventually generates a list of combinations ascendingly ordered by their execution time.
\end{enumerate}
It is worth noting that we designed the above tuple in a configurable way, which means that we can always revise the tuple (e.g. adding or removing items, modifying the candidate values of an item) as needed.

Empirically, the local search of a CNN model takes a few hours using one machine, which is acceptable as it is one-time work. For example, it took about 6 hours to search for the 20 different \emph{CONV} workloads of ResNet-50 on an 18-core Intel Skylake processor. In addition, we can maintain a database to store the results for every convolution workload (defined by the feature map and convolution kernel sizes) on every CPU type to prevent repeating search for the same convolution in different models.

Local search works well for each individual operation and indeed finds better optimization scheme than our manual work. However, greedily adopting the local optimal of every operation may not lead to the global optimal. Consider two consecutive \emph{CONV} operations \emph{conv\_0} and \emph{conv\_1}, if the output split factor (\emph{oc\_bn}) of \emph{conv\_0} is different from the input split factor (\emph{ic\_bn}) of \emph{conv\_1}, a \emph{LayoutTransform} node needs to be inserted to the graph as discussed in Section~\ref{sec:opt:graph}. This transformation overhead can be too expensive to take advantage of the benefit brought by the local optimal, especially when the data size of the network is large. On the other hand, if we maintain the same split factor throughout the entire network (as we did in Section~\ref{sec:opt:graph}), we may miss the opportunity to optimize some \emph{CONV}s. Therefore, a trade-off should be made using a global search.

\subsubsection{Global search}
\label{sec:opt:search:global}
In this subsection, we extend the optimization search to the entire computation graph. The idea is to allow each \emph{CONV} freely choosing the split factor $x$ (i.e. $ic\_bn$ and $oc\_bn$), and take the corresponding data layout transformation time into consideration. According to Section~\ref{sec:opt:graph}, the operations between \emph{CONV}s are either \emph{layout-oblivious} or \emph{layout-tolerant}, so they can use whatever $x$ decided by the \emph{CONV} operation. 

We extract a snippet of a typical CNN model in Figure~\ref{fig:search} to illustrate the idea. From the figure we see that each \emph{CONV} has a number of candidate schemes specified by different (\emph{ic\_bn} and \emph{oc\_bn}) pairs. The shortest execution time achieved by each pair can be obtained in the local search step. The number of pairs is bound to 100 since both \emph{ic\_bn} and \emph{oc\_bn} usually have choices less than 10. Choosing different schemes will introduce different data transformation overheads (denoted in dashed boxes between \emph{CONV}s) or no transformation (if the \emph{oc\_bn} of the \emph{CONV} equals the \emph{ic\_bn} of its successor). For simplicity, in the figure we omit the operations which do not impact the global search decision such as \emph{ReLU}, \emph{Batch\_Norm} between two \emph{CONV}s. However, operations like \emph{Elementwise\_Add} could not be omitted since it requires the layout of its two input operands (outputs of $CONV_j$ and $CONV_k$ in the figure) to be the same.

Naively speaking, if a CNN model consists of $n$ \emph{CONV}s, each of which has $k_i$ candidate schemes, the total number of options of the global scheme will be $\prod_{i=1}^{n} k_{i}$, very easy to become intractable as the number of layers $n$ grows. Fortunately, in practice, we can use a dynamic programming (DP) algorithm to efficiently solve this problem. Note that when choosing the scheme for a \emph{CONV}, we only need to consider the data layout of it and its direct predecessor(s) but not any other ancestor \emph{CONV}s as long as the so-far globally optimal schemes up to the predecessor(s) are memorized.

Therefore, a straightforward algorithm is constructed in Algorithm~\ref{alg:dp}. In practice, a lot of CNN models has the structure as simple as a list, in which each \emph{CONV} only has one predecessor~\cite{krizhevsky2012imagenet,simonyan2014very}. In this case, after a \emph{CONV} is done, the intermediate states stored for its predecessor can be safely removed. For networks with more complex structure like using \emph{Elementwise\_Add} to add two \emph{CONV} outputs to feed to the next \emph{CONV}~\cite{he2016deep}, it is trickier since the schemes of a \emph{CONV} may need to be saved for a future use (e.g. in Figure~\ref{fig:search} $CONV_l$ needs the schemes of $CONV_j$ via \emph{Elementwise\_Add}).

\begin{algorithm}[htbp]
	\caption{Global search algorithm}
	\label{alg:dp}
	\begin{algorithmic}[1]
		\State Sort the nodes of the graph in topological order
		\State Initialize the optimal schemes of the \emph{CONV}s without dependency using the execution time of their candidate schemes
        \For {$CONV_i$ in topological order}
		\For {each candidate scheme $CSI_j$ of $CONV_i$} \Comment{$j$ is the $j^{th}$ scheme of $CONV_i$}
        \State $t = execution\_time(CSI_j)$
        \State $GSI_j=MAX$ \Comment{initialize global optimal scheme of $CONV_i$ under scheme $j$}
        \For {each so-far globally optimal scheme $GSX_k$ of predecessor $x$} \Comment{$k$ is the $k^{th}$ scheme of $CONV_x$}
        \State $cur\_opt = t+transform\_time(k,j)+GSX_k$
        \If {$cur\_opt < GSI_j$}
        \State $GSI_j = cur\_opt$
        \EndIf
        \EndFor
        \EndFor
        \EndFor \\
        \Return last node's shortest scheme
	\end{algorithmic}
\end{algorithm}

However, if the model structure becomes too complicated with many data dependency links between \emph{CONV}s, the straightforward DP algorithm could go intractable, too. For example, in the object detection model SSD~\cite{liu2016ssd}, the number of states can reach the order of trillions due to the occurrence of many concatenation blocks. In this case, we introduced an approximate solution to accelerate the search. Particularly, we reduced our global search problem to the register allocation problem in the canonical compiler domain with minor modification as follows. The register allocation problem is modeled as graph representation in which each node (variable) has a candidate list containing all possible register options, and each edge is associated with a cost matrix indicating the availability of registers between two nodes~\cite{registerallocpbqp}. Similarly in our global search, each \emph{CONV} has a list of candidate schemes and each edge is associated with the layout transformation cost matrix generated by the scheme lists of two \emph{CONV}s. For other non-\emph{CONV} nodes like \emph{Elementwise\_Add} which require all inputs in the same layout, we fixed the layout of one input and convert all other input layouts to it. Therefore, we defined the candidate list of a non-\emph{CONV} node to be the same as the first input \emph{CONV} and the cost matrix on the edge between these two nodes as all diagonal elements being 0 and all the other elements being infinite. For the edges between this non-\emph{CONV} node and other input nodes, cost matrices are generated from the first input node and other input nodes. After such modification, all nodes and edges in our graph have the valid properties which are required by the register allocation modeling. This enables us to apply a heuristic solver based on partitioned boolean quadratic programming (PBQP) to our problem as it is done in register allocation~\cite{registerallocpbqp}.

In order to verify the result of this approximation algorithm, we compared it with the result of DP (the guaranteed best) on some simple networks where DP is tractable. It turns out that the approximation algorithm gets at least 88\% of the best available result. Empirically, a typical DP search completes in 1 minute for most CNN models. In practice, we switch to the approximation algorithm if DP does not complete in 5 minutes. The approximation algorithm completes quickly, e.g. in 10 seconds. For the 15 popular networks we evaluated in Section~\ref{sec:eval}, only SSD was done approximately.
\section{Evaluation}
\label{sec:eval}
This section evaluates the performance of our proposed solution, \emph{NeoCPU}, by answering the following questions:
\begin{enumerate}
\item What is the overall performance of \emph{NeoCPU} comparing with the start-of-the-art alternatives on various kinds of CPUs?
\item What is the individual contribution of each optimization idea we proposed?
\end{enumerate}
All experiments were done on Amazon EC2 instances. We evaluated \emph{NeoCPU} on three kinds of CPUs, Intel Skylake (C5.9xlarge, 18 physical cores, featured with AVX-512), AMD EPYC (M5a.12xlarge, 24 physical cores, featured with AVX2) and ARM Cortex A72 (A1.4xlarge, 16 physical cores, featured with NEON). Although testing on the cloud, our results of ARM CPUs apply to the ones at the edge devices such as Raspberry Pi and Amazon Echo Dot due to the same architecture. All cores have uniformed memory access.

\emph{NeoCPU} was built on top of the code base of the TVM stack 0.4.0. For CPUs with x86 architecture, we chose two framework-specific solutions and one framework-agnostic solution as baselines for comparison. For the framework-specific solution, we investigated a wide range of options and figured out that MXNet 1.3.1 with Intel MKL-DNN v0.15 enabled has the widest model coverage with the best inference performance compared to others (e.g. Intel Caffe). In addition, we chose TensorFlow 1.12.0 with ngraph v0.12.0-rc0 integration (empirically proved to be better than TensorFlow XLA on CPUs) due to its popularity. TensorFlow is known to have better performance on CPUs than another popular deep learning framework PyTorch~\cite{coleman2017dawnbench}. The latest Intel OpenVINO Toolkit 2018 R5.445 served as the framework-agnostic solution. We used the official image-classification sample~\footnote{\url{https://docs.openvinotoolkit.org/latest/_inference_engine_samples_classification_sample_README.html}} and object-detection-ssd sample~\footnote{\url{https://docs.openvinotoolkit.org/latest/_inference_engine_samples_object_detection_sample_ssd_README.html}} for benchmarking. For ARM CPUs, we chose MXNet 1.3.1 with OpenBlas 0.2.18 and TensorFlow 1.12.0 with Eigen fd68453~\footnote{\url{https://github.com/tensorflow/tensorflow/blob/r1.12/tensorflow/workspace.bzl\#L128}} as the baselines. No framework-agnostic comparison was performed as on ARM CPUs there is no counterpart of OpenVINO to x86 CPUs. In addition, OpenMP 4.5 implemented in GCC 7.3 was used in the comparison with our own thread pool for multi-thread scalability. As a note, all implementations used direct convolution. Incorporating the advanced convolution algorithms to further improve the performance remains for future work.

We ran the model inference on a number of popular CNN models, including ResNet~\cite{he2016deep}, VGG~\cite{simonyan2014very}, DenseNet~\cite{huang2017densely}, Inception-v3~\cite{szegedy2016rethinking}, and SSD~\cite{liu2016ssd} using ResNet-50 as the base network. Models consumed by MXNet and OpenVINO were from the Gluon Model Zoo~\footnote{\url{https://mxnet.incubator.apache.org/api/python/gluon/model_zoo.html}}. Models consumed by TensorFlow were obtained mostly from TF-SLim~\footnote{\url{https://github.com/tensorflow/tensorflow/tree/master/tensorflow/contrib/slim}} and for some missing ones (e.g. ResNet-34, DenseNet-169) we manually created them. The same model in different formats are semantically identical. As inherited from the TVM stack, \emph{NeoCPU} is compatible to both Gluon and TF-slim formats, and in the evaluation we used the former one. The input data of the model inference are $224 \times 224$ images, except for the Inception Net ($299 \times 299$) and SSD ($512 \times 512$) by following the popular convention. Since the most important performance criterion of model inference is the \emph{latency}, we did all experiments with batch size 1, i.e. each time only one image was fed to the model, to measure the inference time. Therefore, we fix the value $N$ in $NCHW[x]c$ as 1. \emph{NeoCPU} works for larger batch sizes as well, in which cases we just need to add the $N$ value to our configuration tuple.

Since our optimization does not change the semantics of the model, we do not expect any change of the model output. As a sanity check, we compared the results generated by \emph{NeoCPU} with other baselines (prediction accuracy for image classification models and mean accuracy prediction for object detection models) to validate the correctness.

\subsection{Overall Performance}
\label{sec:eval:overall}
We first report the overall performance we got for 15 popular CNN models comparing with the baselines on different CPUs in Table~\ref{tbl:overall}. The results were obtained by averaging the execution times of 1000 samples, doing inference for one at a time. In general, \emph{NeoCPU} is more efficient across different models on different CPU architectures than any of the baselines (up to 11$\times$ speedup without considering the suspicious OpenVINO outliers which will be explained later). Compared to the \emph{best available} baseline result for each model, \emph{NeoCPU} gets 0.94-1.15$\times$ performance on the Intel Skylake CPU, 0.92-1.72$\times$ performance on the AMD EYPC CPU, and 2.05-3.45$\times$ performance on the ARM Cortex A72 CPU.

\begin{table*}[t]
\begin{subtable}{\textwidth}
\centering
	\resizebox{\textwidth}{!}{%
        \begin{tabular}{ccccccccc}
        \toprule
        Unit: ms & ResNet-18 & ResNet-34 & ResNet-50 & ResNet-101 & ResNet-152 & VGG-11 & VGG-13 & VGG-16 \\ \hline \\[-1.75ex]
        MXNet & 2.77, .01 & \textbf{4.85, .02} & 6.60, .00 & 12.90, .04 & 18.58, .07 & 12.05, .00 & 15.16, .00 & 18.55, .00 \\
        TensorFlow & 4.07, .00 & 6.95, .00 & 11.93, .01 & 20.36, .00 & 37.33, .02 & 18.78, .01 & 24.28, .00 & 27.64, .02 \\
        OpenVINO & 3.54, .00 & 5.43, .00 & 7.95, .00 & 12.55, .00 & 17.32, .01 & 138.07, .12 & 137.51, .14 & 140.95, .33 \\
        NeoCPU & \textbf{2.64, .00} & 5.14, .00 & \textbf{5.73, .00} & \textbf{11.15, .01} & \textbf{17.24, .01} & \textbf{11.91, .00} & \textbf{14.91, .00} & \textbf{18.21, .00}          \\
        \toprule
        & VGG-19 & DenseNet-121 & DenseNet-161 & DenseNet-169 & DenseNet-201 & Inception-v3 & SSD-ResNet-50 & \\ \hline \\[-1.75ex]
        MXNet & 21.83, .00 & 14.72, .00 & 31.07, .01 & 19.73, .00 & 26.66, .00 & \textbf{10.43, .00} & 42.71, .00 & \\
        TensorFlow & 35.94, .00 & 18.65, .01 & 32.97, .00 & 23.03, .01 & 29.19, .01 & 16.39, .04 & 358.98, .13 & \\
        OpenVINO & 147.41, .12 & 9.03, .00 & 18.55, .01 & 11.80, .01 & 14.92, .01 & 10.65, .00 & 30.25*, .01 & \\
        NeoCPU & \textbf{21.77, .00} & \textbf{8.04, .01} & \textbf{17.45, .04} & \textbf{11.21, .01} & \textbf{13.97, .03} & 10.67, .01 & \textbf{31.48, .00} & \\ \hline
        \end{tabular}
    }
\caption{Overall performance on a system with 18-core Intel Skylake CPU}
\label{tbl:intel}
\end{subtable}

\begin{subtable}{\textwidth}
\centering
	\resizebox{\textwidth}{!}{%
        \begin{tabular}{ccccccccc}
        \toprule
            Unit: ms & ResNet-18 & ResNet-34 & ResNet-50 & ResNet-101 & ResNet-152 & VGG-11 & VGG-13 & VGG-16 \\ \hline \\[-1.75ex]
            MXNet & 7.84, .36 & 14.66, .14 & 22.48, .48 & 40.57, 2.54 & 58.92, 3.21 & 49.17, 1.75 & 59.19, 1.35 & 72.57, 2.74 \\
            TensorFlow & 13.95, .24 & 25.02, .49 & 38.14, .35 & 74.41, .56 & 108.38, .24 & 60.30, .22 & 71.16, .33 & 96.33, .22 \\
            OpenVINO & 8.56, 1.02 & 15.18, .60 & 21.95, .42 & 1711.42, 1.59 & 2515.08, 2.51 & 662.09, 1.73 & 709.58, 1.78 & 828.17, 2.09 \\
            NeoCPU & \textbf{7.15, .49}  & \textbf{14.10, .68} & \textbf{18.79, 1.01} & \textbf{39.32, .87} & \textbf{55.71, .54} & \textbf{28.58, .74} & \textbf{38.17, .29} & \textbf{57.63, .68} \\
        \toprule
             & VGG-19 & DenseNet-121 & DenseNet-161 & DenseNet-169 & DenseNet-201 & Inception-v3 & SSD-ResNet-50 & \\ \hline \\[-1.75ex]
            MXNet & 84.76, 1.91 & 35.00, 1.06 & 79.58, .63 & 47.82, 1.67 & 63.67, .15 & 30.12, .09 & 132.73, 2.59 & \\
            TensorFlow & 121.04, .38 & 45.87, .15 & 98.39, .93 & 57.49, .28 & 77.37, .24 & 48.78, .45 & 747.78, 2.24  & \\
            OpenVINO & 1113.17, 2.39 & \textbf{22.36, .24} & 818.86, 1.39 & 438.72, 1.27 & 453.12, 1.75 & \textbf{25.75, .83} & 93.65*, .81 & \\
            NeoCPU & \textbf{63.78, .18} & 24.30, .54 & \textbf{49.37, .09} & \textbf{31.70, .47} & \textbf{46.12, .51} & 26.37, .32 & \textbf{97.26, .54} & \\ \hline
        \end{tabular}
    }
\caption{Overall performance on a system with 24-core AMD EYPC CPU}
\label{tbl:amd}
\end{subtable}

\begin{subtable}{\textwidth}
\centering
	\resizebox{\textwidth}{!}{%
        \begin{tabular}{ccccccccc}
        \toprule
            Unit: ms & ResNet-18 & ResNet-34 & ResNet-50 & ResNet-101 & ResNet-152 & VGG-11 & VGG-13 & VGG-16 \\ \hline \\[-1.75ex]
            MXNet & 75.82, 1.31 & 135.24, 2.49 & 149.65, 2.37 & 252.76, 3.25 & 351.60, 3.49 & 385.50, 2.39 & 505.06, 3.28 & 575.80, 2.98 \\
            TensorFlow & 50.50, .07 & 96.50, .11 & 107.50, .12 & 223.83, .17 & 336.56, .19 & 245.97, .18 & 336.05, .27 & 381.46, .21 \\
            NeoCPU & \textbf{19.26, .08} & \textbf{37.20, .14} & \textbf{45.73, .02} & \textbf{86.77, .08} & \textbf{126.65, .13} & \textbf{87.66, .21} & \textbf{124.75, .05} & \textbf{162.49, .14} \\
        \toprule
             & VGG-19 & DenseNet-121 & DenseNet-161 & DenseNet-169 & DenseNet-201 & Inception-v3 & SSD-ResNet-50 & \\ \hline \\[-1.75ex]
            MXNet & 642.27, 4.30 & 211.54, 3.22 & 389.33, 2.98 & 264.36, 3.82 & 315.10, 3.49 & 275.28, 3.27 & 657.22, 3.29 & \\
            TensorFlow & 459.91, .27 & 122.48, .07 & 301.51, .11 & 159.39, .08 & 204.79, .10 & 142.00, .07 & 1020.16, .47 & \\
            NeoCPU & \textbf{201.03, .49} & \textbf{44.00, .09} & \textbf{87.36, .15} & \textbf{58.93, .65} & \textbf{65.48, .54}  & \textbf{84.00, .08} & \textbf{318.48, .11} & \\ \hline
        \end{tabular}
    }
\caption{Overall performance on a system with 16-core ARM Cortex A72 CPU}
\label{tbl:arm}
\end{subtable}
\caption{Overall performance of \emph{NeoCPU} and the selected baselines. Each entry contains the mean value of 1000 runs and the corresponding standard error. The best performance of each model is in \textbf{bold}. (*OpenVINO on Intel and AMD CPUs does not measure the entire SSD execution time)}
\label{tbl:overall}
\end{table*}

As framework-specific solutions, MXNet and TensorFlow were suboptimal for CNN inference on CPUs because it is lacking of flexibility to perform sufficient graph level optimization (e.g. flexible data layout management). MXNet has active MKL-DNN support from Intel so it performed quite well on CPUs with the x86 architecture. MXNet performed worse than TensorFlow on ARM due to the scalability issue (demonstrated in Figure~\ref{fig:scalability:inception}). TensorFlow performs significantly worse on SSD as it introduces branches to this model, which requires dynamic decisions to be made during the runtime. Comparatively, the framework-agnostic solution provided by the OpenVINO tries to further boost the performance by removing the framework limitation. However, the performance of OpenVINO was unstable across models. Although it gets appealing results on some cases, OpenVINO sometimes performed extremely slowly on certain models (e.g. 45$\times$ slower than us for ResNet-152 on AMD) for unknown reasons. When summarizing the speedup results, we do not include these outliers. It is also worth noting that the OpenVINO measures the execution time of SSD without taking into account a significant amount of operations including \emph{multibox detection}. Since OpenVINO is not open-sourced, we were not able to modify it for apples-to-apples comparison on the SSD model. OpenVINO does not work for ARM CPUs as it relies on MKL-DNN which optimizes only for CPUs with x86 architecture. \emph{NeoCPU} outperforms the baselines mostly because of the advanced optimization techniques we presented in Section~\ref{sec:opt}. In addition, all baselines largely rely on the third-party libraries (MKL-DNN, OpenBlas, Eigen) to achieve good performance. \emph{NeoCPU}, on the other hand, is independent from those high-performance libraries, which gives us more room to optimize the model inference as a whole.

\subsection{Optimization Implications}
\label{sec:eval:opt}
This subsection breaks up the end-to-end performance gain of \emph{NeoCPU} by investigating the performance boost of each individual optimization technique we described in Section~\ref{sec:opt}. For the sake of space, in each comparison we only pick one network from a network family, respectively. Other networks in the same family share the similar benefits. We only report the performance results on Intel CPUs in Section~\ref{sec:eval:layout}-\ref{sec:eval:search}. The optimization effect applies to AMD and ARM CPUs, too. Basically, Section~\ref{sec:eval:layout} is the operation-level optimization, and Section~\ref{sec:eval:elim} and~\ref{sec:eval:search} cover the operation- and graph-level joint optimization.
\subsubsection{Layout optimization of \emph{CONV}}
\label{sec:eval:layout}
Firstly, we compare the performance with and without organizing the data in a memory access and vectorized instruction utilization friendly layout (\emph{NCHW\{x\}c}) for the \emph{CONV} operations at the second row of Table~\ref{tbl:speedup}. This is the operation-level optimization that is commonly applied by the compared baselines in Section~\ref{sec:eval:overall}. We replicate it as a template using TVM scheduling schemes without touching the assembly code or intrinsics, which enables the subsequent optimization for various CNN models on different CPU architectures. From row 2 of Table~\ref{tbl:speedup} we see significant improvement compared to the default data layout (\emph{NCHW}), whose performance is normalized to baseline 1. Both implementations are with proper vectorization and thread-level parallelization, as well as basic graph-level optimizations introduced by the original TVM stack, e.g. operation fusion, pre-computing, inference simplification, etc.

\begin{table}[tbp]
	\centering
	\resizebox{0.5\textwidth}{!}{%
		\begin{tabular}{l c c c c c}
			\midrule
			\textbf{Speedup} & \textbf{ResNet-50} & \textbf{VGG-19} & \textbf{DenseNet-201} & \textbf{Inception-v3} & \textbf{SSD-ResNet-50} \\
			\hline\\[-1.75ex]
            Baseline & 1 & 1 & 1 & 1 & 1\\
            Layout Opt. & 5.34 & 8.33 & 4.08 & 7.41 & 6.34 \\
            Transform Elim. & 8.22 & 9.33 & 5.51 & 9.11 & 9.32 \\
			Global Search & 12.25 & 10.54 & 6.89 & 11.85 & 12.49 \\
			\midrule
		\end{tabular}
	}
	\caption{The individual speedup brought by our optimization compared to the \emph{NCHW} baseline. The speedup of row $n$ was achieved by applying the optimization techniques till this row.}
	\label{tbl:speedup}
\end{table}

\subsubsection{Layout transformation elimination}
\label{sec:eval:elim}
Secondly, we evaluate the performance boost brought by eliminating the data layout transformation overhead as discussed in Section~\ref{sec:opt:graph}. The results were summarized at the third row of Table~\ref{tbl:speedup}. Compared to the layout optimization of \emph{CONV} (second row of Table~\ref{tbl:speedup}), layout transformation elimination further accelerates the execution time by $1.1 - 1.5\times$. \emph{NeoCPU} uses a systematic way to eliminate the unnecessary data layout transformation by inferring the data layout throughout the computation graph and inserting the layout transformation nodes only if needed, which is not seen in other works.

\subsubsection{Optimization scheme search}
\label{sec:eval:search}
Next, we compare the performance between the optimization schemes produced by our search algorithm and the ones carefully picked by us manually. By comparing the third and fourth row of Table~\ref{tbl:speedup}, our algorithm (described in Section~\ref{sec:opt:search}) is able to find the (approximately) best combination of data layouts which outperforms the manually picked results by $1.1 - 1.5\times$. ResNet-50 (and its variants) gains more speedup from global search because the network structure is more complicated, hence leaving more optimization room. In contrast, VGG-19 (and its variants) gains less since the structure of this model is relatively simple. SSD utilizes the approximation algorithm and gets significant speedup, too. The results also verify that, with automatic search, we can get rid of the tedious manual picking of parameters by producing even better results. To the best of our knowledge, \emph{NeoCPU} is the only one that does this level of optimization.

\subsubsection{Multi-thread parallelization}
\label{sec:eval:scale}
\begin{figure}[!ht]
	\centering
	\begin{subfigure}{0.47\textwidth}
		\centering
		\includegraphics[width=\textwidth]{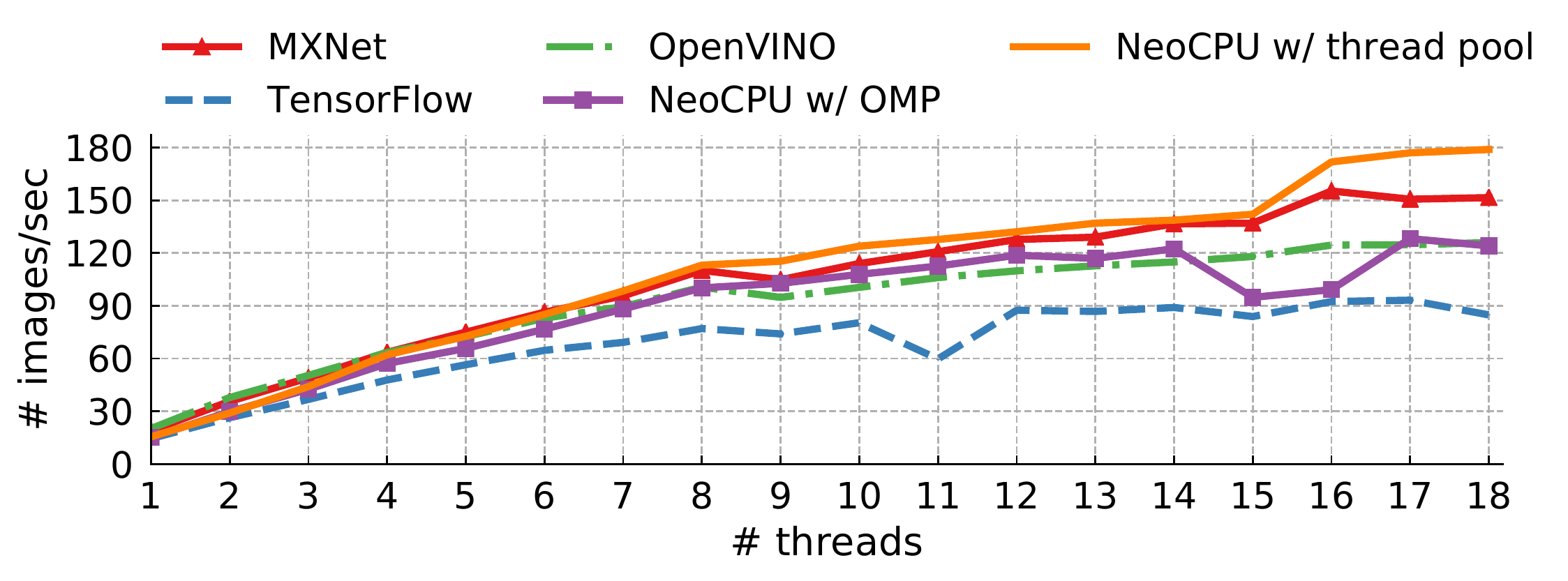}
		\caption{ResNet-50 on a system with 18-core Intel Skylake CPU}
		\label{fig:scalability:resnet}
	\end{subfigure}
	\\
	\begin{subfigure}{0.47\textwidth}
		\centering
		\includegraphics[width=\textwidth]{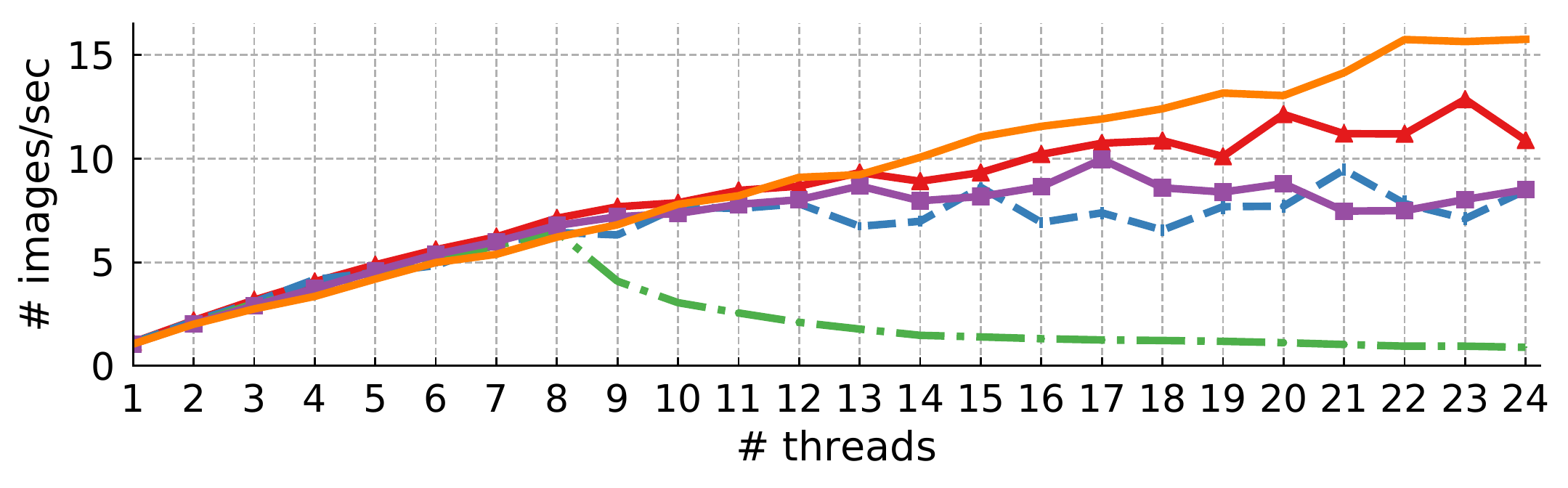}
		\caption{VGG-19 on a system with 24-core AMD EPYC CPU}
		\label{fig:scalability:vgg}
	\end{subfigure}
	\\
    \begin{subfigure}{0.47\textwidth}
		\centering
		\includegraphics[width=\textwidth]{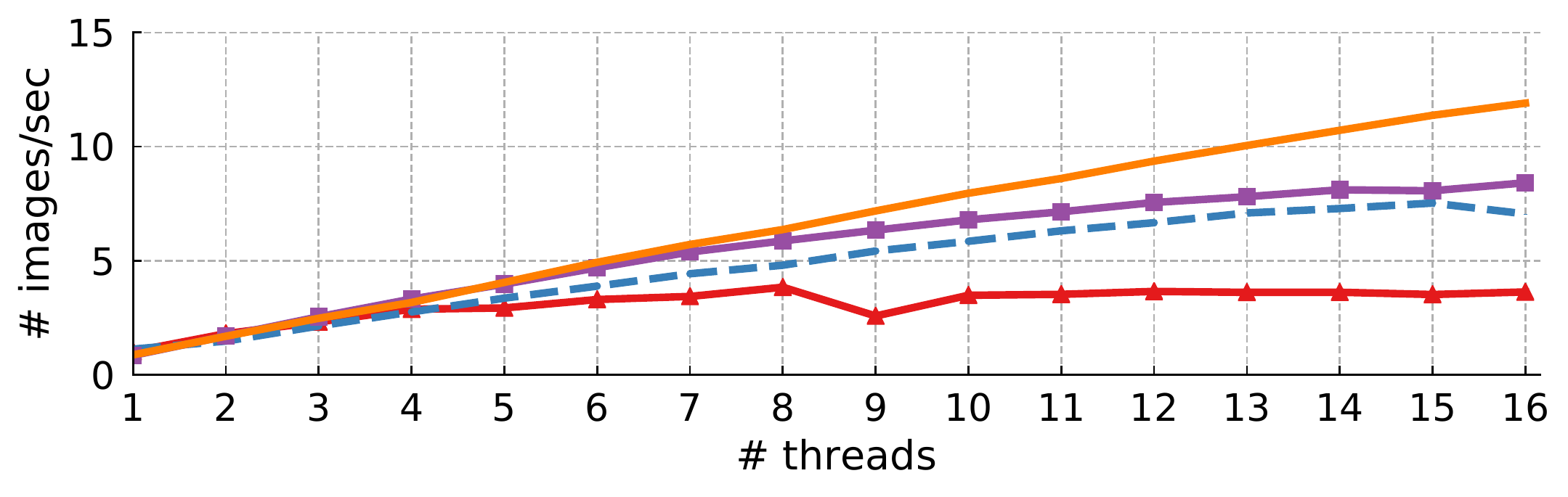}
		\caption{Inception-v3 on a system with 16-core ARM Cortex A72 CPU}
		\label{fig:scalability:inception}
	\end{subfigure}
	\caption{Scalability comparison between different multi-threading implementations. The standard errors ($<0.4$) are too small to be visible in the diagrams.}
	\vspace{-2em}
	\label{fig:scalability}
\end{figure}

Lastly, we did a strong scalability experiment using the multi-threading implementations backed by our own thread pool described at Section~\ref{sec:opt:op:parallel} and the commonly used OpenMP API implemented in the GCC compiler. We also included the result of MXNet, TensorFlow and OpenVINO using Intel MKL-DNN, OpenBlas or Eigen (all realizing multi-threading via OpenMP) for comparison. We configured OpenMP via environment variables to make sure that the jobs are statically partitioned and each thread runs on a disjoint core, which resemble the behavior of our thread pool for apples-to-apples comparison. Figure~\ref{fig:scalability} summarizes the number of images a model can inference one by one (i.e. $batch\ size = 1$) in a second as a function of the number of threads the model inference uses. For the sake of space, we demonstrate one result for one CPU type. The figure shows that our thread pool achieves better scalability than OpenMP in \emph{NeoCPU} as well as in the baselines. Although the tasks are embarrassingly parallelizable, each model inference consists of a number of parallelization regions. The overhead of OpenMP to launch and suppress threads before and after a region is larger than our thread pool, which attributes to the less scalability of OpenMP. Furthermore, sometimes we observed that the performance obtained by OpenMP jitters, or even drops, while adding threads. In addition, the performance of OpenMP may differ across different implementations. In summary, our evaluation suggests that in our use cases, it is preferable to have a self-customized thread pool with full control.

\section{Related Works}
\label{sec:related}
As deep learning demonstrates more and more power in the real-world applications, there is a significant amount of effort being made to accelerate the deep learning workloads on all kinds of hardware ranging from CPUs~\cite{zlateski2016znn,heinecke2016libxsmm,mkldnn,2018arXiv180709667T}, GPUs~\cite{chetlur2014cudnn,chen2014efficient}, FPGAs~\cite{farabet2009cnp,zhang2015optimizing,han2017ese}, to special-purpose accelerators~\cite{chen2016diannao,Jouppi2017TPU}. Modern deep learning frameworks normally leverage these optimized implementations to run deep learning training and inference on the corresponding hardware targets. There are also works tailored for inference to address the inference-specific requirement such as low latency and small binary size on different hardware targets (e.g. GPUs~\cite{tensorrt}, ASICs~\cite{han2017ese}). \emph{NeoCPU} is more flexible and combines the operation- and graph-level optimization intelligently. Although this paper focuses on CPUs, the ideas are applicable to other hardware targets. 

\emph{NeoCPU} is based on the TVM stack~\cite{chen2018tvm}, an end-to-end framework inspired by Halide~\cite{Ragan-Kelley2013Halide}, which expresses a deep learning model into intermediate representations (IRs) and compiles to the machine code. There are several other similar deep learning compilers such as TensorFlow XLA~\cite{leary2017xla}, Tensor Comprehensions~\cite{vasilache2018tensor}, Glow~\cite{rotem2018glow} and DLVM~\cite{wei2017dlvm}. However, so far none of them has reported CPU inference results on par with what we did (e.g. Glow only optimized single-core performance on CPUs). We believe our proposed solution could be an integral part to these frameworks.

We follow the well-studied ideas implemented in other high-performance libraries~\cite{mkldnn,xianyi2014openblas} to optimize the computationally-intensive \emph{CONV} operations. In addition to the libraries, there are also highly customized optimization works for convolutions and matrix multiplications on Intel CPUs~\cite{georganas2018anatomy,heinecke2016libxsmm}. These works are mostly about individual operation-level optimizations, which do not consider maintaining data layouts through the entire network. Specifically, they carefully investigate the computation nature of convolutions as well as the available CPU resources to fine tune the operations. This kind of optimization is able to maximize the convolution performance on the targeted CPUs but is not very flexible to extend to other platforms and to do joint optimization. Unlike others, we make the optimization as a configurable template so that it is flexible to fit to different CPU architectures and enable the opportunity to surpass manually tuned performance via operation- and graph-level joint optimization.

Our work utilizes auto search to look for optimal solutions. Similar \emph{auto-tuning} ideas were used in other works as well~\cite{whaley1998automatically,vasilache2018tensor,chen2018learning}. However, they all focused on performance tuning for single operations, while ours extends the scope to the entire CNN model to search for optimal solutions globally. Recently, we also observed other work optimizing the DNN workloads at the graph level~\cite{jiaoptimizing}. This work attempts to obtain better global performance using relaxed graph substitutions which may harm the local performance within a few operations. Its non-greedy search idea is conceptually similar to ours and potentially applicable to our solution. The approximation algorithm we employed to deal with the global search for the models with complicated structures (e.g. SSD) is inspired by the application of PBQP in the register allocation problem~\cite{registerallocpbqp,colorregister,eckstein2003code}. This paper leverages the previous idea and applies to a new domain by minor modification.
\section{Conclusion}
\label{sec:concl}
In this paper, we proposed an end-to-end solution to compile and optimize convolutional neural networks for efficient model inference on modern CPUs. The experiments show that we are able to achieve up to $3.45\times$ speedup on 15 popular CNN models on the various kinds of CPUs (Intel Skylake, AMD EPYC and ARM Cortex A72) compared to the performance of the state-of-the-art solutions. The future work includes extending to other convolution computation algorithms such as Winograd and FFT, handling model inference in quantized values (e.g. INT8) and extending our operation- and graph-level joint optimization ideas to work on other hardware platforms (e.g. NVidia GPUs compared with TensorRT). Supporting the optimized model inference in dynamic shapes (e.g. RNNs~\cite{holmes2019grnn,zhang2018deepcpu}) is another interesting direction to explore.

\section*{Acknowledgments}
We would like to thank our shepherd Peter Pietzuch and
the anonymous reviewers of the USENIX ATC program committee for
their valuable comments which improved the paper a lot. We
are also grateful to Tianqi Chen and Animesh Jain for helpful discussion and constructive suggestion.

\bibliographystyle{plain}
\bibliography{ref}
\balance


\end{document}